\documentstyle[aps,multicol,graphics,prb,times]{revtex}

\begin{document}
\draft
\title{Relaxational dynamics study of the classical Heisenberg spin
$XY$ model in spherical coordinate representation}
\author {Beom Jun Kim and Petter Minnhagen}
\address {Department of Theoretical Physics, Ume{\aa} University,
901 87 Ume{\aa}, Sweden}
\author{Suhk Kun Oh and Jean S. Chung}
\address {Department of Physics, Chungbuk National University, Cheongju 361-763, Korea}
\maketitle
\begin{abstract}
The two- and three-dimensional classical Heisenberg spin $XY$ (CHS$XY$) models, 
with the spherical coordinates of spins taken as dynamic variables,
are numerically investigated. We allow the polar $\theta$ and 
azimuthal $\phi$ angles to have uniform values in $[0,\pi)$
and $[-\pi, \pi)$, respectively, and the static universality class 
is shown to be identical to the classical $XY$ model with two-component spins,
as well as the CHS$XY$ model with a different choice of dynamic 
variables, conventionally used in the literature.
The relaxational dynamic simulation reveals that
the dynamic critical exponent $z$ is found to
have the value $z \approx 2.0$ for both two and three dimensions, in contrast
to $z \approx d/2$ ($d=$ spatial dimension) found previously with spin dynamics
simulation of the conventional CHS$XY$ model. 
Comparisons with the usual two-component classical $XY$ model are also made.

\end{abstract}

\pacs{PACS numbers: 75.40.Gb, 75.40.Mg, 64.60.Ht}
\begin{multicols}{2}

\section{Introduction}
The static critical behaviors of the $XY$ model in
two and three dimensions (2D and 3D) have been studied for more than
20 years and there exists well-established consensus on
the nature of the phase transitions and the values of
the critical exponents.~\cite{KT,minnhagen:rmp} For example, static universal properties
have been well established where the static exponents
do not depend on details of models.
On the other hand,
the dynamic universality class has still not been completely sorted out.~\cite{minnhagen:rmp,ahns}

The usual $XY$ model, where the spins are two dimensional,
has been found to have dynamic critical
exponents $z$ that seem to depend both on the dynamic model used
and on the quantity measured.\cite{bjkim3dmc}
In 2D most of the existing works have obtained the result that
$z \approx 2.0$ at the Kosterlitz-Thouless (KT) transition~\cite{KT} 
temperature $T_{KT}$ and that $z$ increases as the temperature 
$T$ is lowered below $T_{KT}$.~\cite{minnhagen:rmp,ahns,bjkim01,bjkim99a,larsB,holmlund96,weber2D} 
However, the result that $z \approx 2.0$ in the
whole low-temperature phase has also been found.~\cite{shortXY,skkim}
In 3D, on the other hand, there is a growing consensus that the dynamic
critical exponent $z$ associated with voltage (or phase slip)
fluctuations is $z \approx 1.5$ (Refs.~\onlinecite{larsB,lars3D,LCG3D,khlee92}) 
although a rigorous analytic justification is still 
lacking. Furthermore this appears to be the case even for relaxational
dynamics in spite of the fact that  $z \approx 2.0$ has been concluded from the standard
dynamic renormalization group method (for example,
in Ref.~\onlinecite{wickham00}) in accordance with model A in the
Hohenberg-Halperin classification.\cite{hh}

The variant of the $XY$ model, which we study here, is given by a
Hamiltonian of the same form as the usual $XY$ model but where the spins
are three dimensional [we call this the classical Heisenberg
spin $XY$ model (CHS$XY$) to avoid confusion with the usual
$XY$ model]. This model has previously been studied
subject to the so-called ``spin dynamics.''\cite{SD2D,skoh,krech99}
Although the CHS$XY$ model belongs to the same
static universality class as the $XY$ model both in
2D (Ref.~\onlinecite{cuccoli95}) and 3D (Ref.~\onlinecite{nho99}),
studies of spin dynamics for the CHS$XY$ model 
have given $z \approx 1.0$ in 2D,~\cite{SD2D}
which differs significantly from $z \approx 2.0$ in the $XY$ model.
In 3D, while Ref.~\onlinecite{skoh}
has found $z \approx 1.5$, the possibility 
of a breakdown of the dynamic scaling has been suggested, i.e., 
that $z$ is not unique but
has different values $z_x = 1.38(5)$ and $z_z = 1.62(5)$ 
for the decay of correlations in the in-plane and the out-of-plane
directions, respectively.~\cite{krech99}

We here propose a variant of the CHS$XY$ model where
the spherical coordinates of the spins are taken as the dynamic
variables with a uniform measure in phase space, and
investigate the dynamic critical behaviors of the model
in two and three dimensions subject to 
relaxational dynamics instead of spin dynamics.
Relaxational dynamics belongs to
the model A with the expected value $z\approx 2$ in the Hohenberg-Halperin 
classification.~\cite{hh} However, this value does not always seem
to be guaranteed.
For example, the purely relaxational form
of dynamics applied to the $XY$ model in 3D under the fluctuating 
twist boundary condition has been found to give $z\approx 1.5$ 
(Ref.~\onlinecite{lars3D}). Even the Monte Carlo (MC) dynamics
simulations, which are generally believed to correspond to relaxational dynamics,  
for the 3D $XY$ model with both phase~\cite{bjkim3dmc} 
and vortex~\cite{LCG3D} representation have also led 
to $z\approx 1.5$. 

The paper is organized as follows. In Sec.~\ref{sec:model},
the Hamiltonian of the CHS$XY$ model in the spherical coordinate
representation and the corresponding equations of motion for the relaxational 
dynamics are introduced.
Although our main interests are in dynamic critical behaviors
we also perform Monte Carlo simulations in Sec.~\ref{sec:mc}
to confirm the equivalence with the conventional CHS$XY$ model and
then compare with static and dynamic results from the relaxational
dynamics in Sec.~\ref{sec:rd},
which constitutes the main results of the current work.
Finally we devote Sec.~\ref{sec:sum} for summary and discussions.

\section{Model} \label{sec:model}
We begin with the Hamiltonian of the conventional CHS$XY$ model in the $d$-dimensional
hypercubic geometry with size $N = L^d$ ($L$ is the linear size),
\begin{equation} \label{eq:Hxy}
H[\{s^x, s^y\}] = -J\sum_{\langle ij\rangle}(s_i^x s_j^x + s_i^y s_j^y),
\end{equation}
where $J$ is the coupling strength, 
the summation is over all nearest-neighbor pairs, and 
the three-dimensional local spin ${\bf s}_i = (s_i^x, s_i^y, s_i^z)$
at site $i$ has unit length ($|{\bf s}_i|^2 = 1$), or equivalently 
the partition function should include the measure
$\delta[(s_i^x)^2 + (s_i^y)^2 + (s_i^z)^2 - 1]$.
The CHS$XY$ model with the Hamiltonian~(\ref{eq:Hxy}) can be viewed as either an
extension from the original $XY$ model where spins are two dimensional, 
or as a special case of the Heisenberg $XXZ$ model with couplings 
only in the $x$-$y$ plane. 

The more convenient representation of the conventional CHS$XY$ Hamiltonian 
is written as
\begin{equation} \label{eq:Hzphi}
H[\{s^z, \phi\}] = -J\sum_{\langle ij\rangle} \sqrt{[1-(s_i^z)^2][1-(s_j^z)^2]}
\cos(\phi_i - \phi_j), 
\end{equation}
where $(s_i^x)^2 + (s_i^y)^2 = 1 - (s_i^z)^2$ has been used, and 
$\phi_i$ is the angle between the $x$-$y$ plane component of the spin ${\bf s}_i$,
i.e., ${\bf s}_i - s_i^z{\hat {\bf z}}$, 
and the positive $x$ axis. In representation~(\ref{eq:Hzphi}), $\phi$ 
and $s^z$ have uniform measure since 
\begin{eqnarray}
& &\int ds^x \int ds^y \int ds^z \delta[ (s^x)^2 + (s^y)^3 + (s^z)^2 - 1] \nonumber \\ 
&=& \int ds^z \int rdr \int d\phi \delta[ r^2 + (s^z)^2 - 1] \nonumber \\
&\propto& \int_{-1}^1 ds_z \int_{-\pi}^\pi d\phi,  
\end{eqnarray}
where $r^2 \equiv (s^x)^2 + (s^y)^2$, $\phi \equiv \arctan(s^y/s^x)$, and the identity
$\delta(r^2 - a^2) = \delta(r - a)/2r$
has been used.

We introduce the polar angle 
variable $\theta$ in the spherical coordinate system as follows:
\begin{eqnarray}
s_i^x & = & \sin\theta_i \cos\phi_i , \nonumber \\
s_i^y & = & \sin\theta_i \sin\phi_i , \nonumber \\
s_i^z & = & \cos\theta_i ,
\end{eqnarray}
which then leads to the representation 
\begin{equation} \label{eq:Htp}
H[\{\theta,\phi\}] = 
-J\sum_{\langle ij\rangle} \sin\theta_i \sin\theta_j \cos(\phi_i - \phi_j).
\end{equation}
We then simplify the conventional CHS$XY$ model and use the {\it uniform}
measure not only for $\phi$ but also for $\theta$ variables.
One advantage of this is that no additional constraint is 
required since $|{\bf s}_i|$ = 1 is satisfied automatically in the 
representation~(\ref{eq:Htp}).
One should note that the conventional CHS$XY$ model [represented by 
the Hamiltonian~(\ref{eq:Hzphi}) with the uniform measure in $s^z$ and $\phi$] 
and its variant model studied in this work [the Hamiltonian~(\ref{eq:Htp})
with the uniform measure in $\theta$ and $\phi$]~\cite{foot:def}
do not have the same partition function and free energy,
and accordingly some nonuniversal
properties like the critical temperature can be different. However,
one expects that universal critical properties should be the same 
as will be clearly confirmed in Sec.~\ref{sec:mc} below.

The relaxational dynamic equations are simply given by~\cite{bjkim99a}
\begin{eqnarray} \label{eq:rd}
\dot\theta_i &=& - \Gamma \frac{\partial H[\{\theta,\phi\}] }{\partial \theta_i} + \eta_i^\theta , \nonumber \\
\dot\phi_i &=& - \Gamma \frac{\partial H[\{\theta,\phi\}] }{\partial \phi_i} + \eta_i^\phi ,
\end{eqnarray}
where $\Gamma$ is a constant that determines the time scale
of relaxation, and the stochastic thermal noise terms satisfy
$\langle \eta_i^\theta(t) \rangle    =  \langle \eta_i^\phi(t) \rangle  
 = \langle \eta_i^\theta(t) \eta_j^\phi(t')  \rangle   =  0$
and
$\langle \eta_i^\theta (t) \eta_j^\theta(t') \rangle  =
\langle \eta_i^\phi (t) \eta_j^\phi(t') \rangle = 2T \delta_{ij} \delta(t-t')$
with the ensemble average $\langle \cdots \rangle$. 
(From now on, we measure the temperature $T$ and the time $t$ in units of 
$J/k_B$ and $1/\Gamma J$, respectively.)
From the Fokker-Planck (FP) formalism, it is straightforward to show
that the stationary solution of the FP equation, corresponding
to the above Langevin-type equations of motion (\ref{eq:rd}), 
is simply the equilibrium Boltzmann distribution with the Hamiltonian 
given in Eq.~(\ref{eq:Htp}). In other words, the relaxational dynamics
used in this work automatically produces equilibrium 
fluctuations in time, which are compatible with the Boltzmann distribution of the 
same Hamiltonian. In this respect, the initial configuration of the
relaxational dynamics can be chosen arbitrarily; the equilibrium 
fluctuations are generated
by the dynamics itself as the system evolves in time. This is in contrast to the
widely used spin dynamics, where the initial configurations
must be generated according to the equilibrium distribution. Otherwise
the spin dynamics will not reflect the properties of the
equilibrium. 
Consequently, the relaxational dynamics described here is consistent
with the usual physical situation of a system
in contact with a thermal reservoir. From this perspective we believe
that the relaxational dynamics can phenomenologically catch 
relevant features for a real spin system in a situation where the thermal
effects are strong.

\section{Monte Carlo Simulation} \label{sec:mc}
For completeness we start by calculating the static properties from
Monte Carlo (MC) simulations within the spherical coordinate
($\{\theta, \phi\}$) representation with both variables uniformly
distributed, which we have not found in the literature.
We use the standard Metropolis algorithm applied to
the Hamiltonian~(\ref{eq:Htp}), and the variations of $\theta_i$ and 
$\phi_i$ at each MC try are tuned to give an acceptance ratio of about
1/2 near the critical temperature. Later we will 
compare the MC results with those 
from the relaxational dynamics in Sec.~\ref{sec:rd}.

\subsection{Three-dimensional lattice} \label{subsec:3dmc}
\begin{figure}
\centering{\resizebox*{!}{6.0cm}{\includegraphics{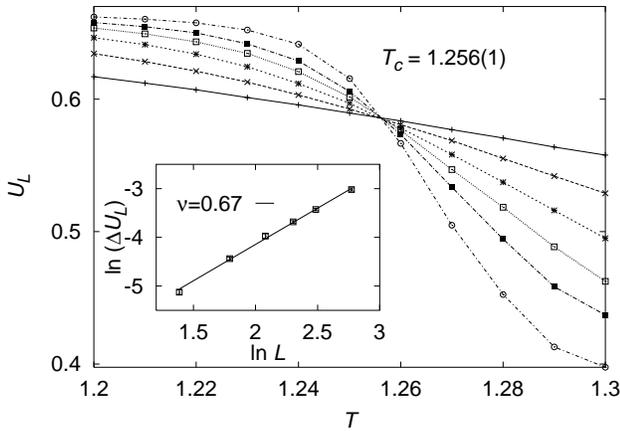}}}
\caption{
Fourth-order Binder's cumulant $U_L$ for the 3D CHS$XY$ model  
in the spherical-coordinate representation 
from MC simulations as a function of the temperature $T$
for various system sizes $L=4$, 6, 8, 10, 12, and 16
(from top to bottom on the right-hand side of the crossing point).
The crossing point gives the estimation of the critical temperature
$T_c = 1.256(1)$.
Inset:
Determination of the critical exponent $\nu$ 
through the finite-size scaling of $U_L$.
$\Delta U_L \equiv U_L(T=1.25) - U_L(T=1.26)$ (see text for details). 
From the least-square fit, $\nu = 0.67(3)$ is obtained.
}
\label{fig:3dmcB}
\end{figure}

In 3D, the transition is detected by the order parameter
defined as~\cite{nho99}
\begin{equation}
\langle m \rangle  \equiv 
\left\langle \frac{1}{N} |{\bf S}| \right\rangle 
 = \left\langle \frac{1}{N} \sqrt{ S_x^2 + S_y^2 + S_z^2 } \right\rangle ,
\end{equation}
where the total spin vector ${\bf S}$ is given by
\begin{equation} \label{eq:totalS}
{\bf S} = \sum_{i=1}^N {\bf s}_i.
\end{equation}
The most convenient way to locate the critical temperature $T_c$ is 
to calculate Binder's fourth-order cumulant 
\begin{equation} \label{eq:Binder}
U_L(T) = 1 - \frac{ \langle m^4 \rangle } { 3 \langle m^2 \rangle^2 },
\end{equation}
which has a unique crossing point as a function of $T$ if plotted
for various system sizes.
Figure~\ref{fig:3dmcB} shows the determination of $T_c$ from
the Binder's cumulant for the system size $L=4$, 6, 8, 10, 12, and 16 
in 3D and $T_c = 1.256(1)$ is obtained. As expected,
this value of $T_c$ is found to be different than $T_c = 1.552(1)$
obtained from the other choice of variables in 
Hamiltonian~(\ref{eq:Hzphi}) with uniform measures for
$s^z$ and $\phi$.~\cite{skoh,krech99,nho99}
However, one expects that such a change cannot 
alter the universality class of the system.
As an example we compute the static critical exponent $\nu$,
which is defined by $\xi \sim (T-T_c)^{-\nu}$ and can be calculated
from 
\begin{equation} \label{eq:Uscale}
U_L(T) \approx U^* + U_1 L^{1/\nu}\left( 1 - \frac{T}{T_c} \right) ,
\end{equation}
where $U^*$ is also a universal value and found to be $U^* = 0.586(1)$
from Fig.~\ref{fig:3dmcB}.
Equation~(\ref{eq:Uscale}) is written in a more convenient form,
\begin{equation}
\Delta U_L = U_L(T_1) - U_L(T_2) \propto L^{1/\nu},
\end{equation}
where $T_1$ and $T_2 (>T_1) $ are picked near $T_c$. In the inset of
Fig.~\ref{fig:3dmcB}, $\Delta U_L$ is plotted as a function of system size $L$ in 
the log scale with $T_1=1.25$ and $T_2=1.26$, and $\nu = 0.67(3)$ is obtained.
The values of $\nu$ and $U^*$  obtained here 
\begin{eqnarray}
\nu &=& 0.67(3) , \\
U^* &=& 0.586(1) , 
\end{eqnarray}
are within error bars in agreement with the known values for the 
conventional CHS$XY$ model with the uniform measure for $s^z$ and $\phi$: 
$\nu = 0.670(7)$, $U^* \approx 0.586$ in Ref.~\onlinecite{nho99}
(the latter was estimated from Fig.~1 in Ref.~\onlinecite{nho99}),
and $\nu = 0.669(6)$, $U^*  = 0.5859(8)$ in Ref.~\onlinecite{krech99}.
This just illustrates that the change of variables and measures in
phase space introduced in Sec.~\ref{sec:model}
does not change the static universality class although the critical
temperature $T_c$ is found to be different.

\subsection{Two-dimensional lattice} \label{subsec:2dmc}
\begin{figure}
\centering{\resizebox*{!}{6.0cm}{\includegraphics{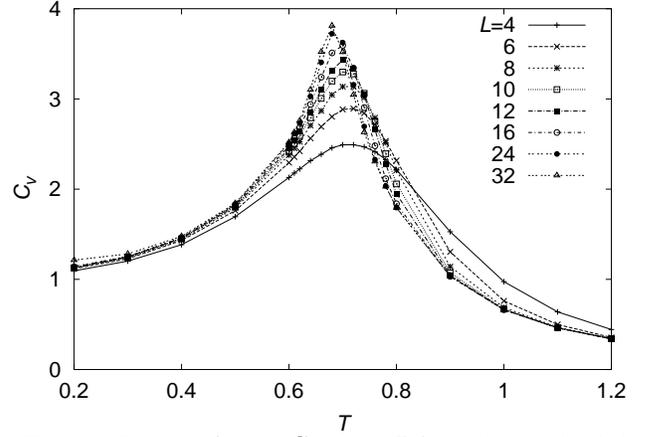}}}
\caption{
The specific heat $C_v$ versus $T$ from the MC simulation of the 2D CHS$XY$ model 
in the spherical-coordinate representation
for various system sizes $L=4$, 6, 8, 10, 12, 16, 24, and 32; 
the specific heat peak appears to saturate as 
$L$ is increased.
}
\label{fig:2dmcCv}
\end{figure}

In many 2D systems with continuous symmetry, the spontaneous magnetization  vanishes
at any nonzero temperature and the phase transition of the $XY$ model
and its related models is of the Kosterlitz-Thouless 
type.~\cite{KT}
Figure~\ref{fig:2dmcCv} shows the specific heat $C_v$, computed from
$C_v = (\langle H^2 \rangle - \langle H \rangle^2)/T^2N$,
for the 2D
CHS$XY$ model in the spherical coordinate representation 
versus $T$ for various system sizes $L=4$, 6, 8, 10, 
12, 16, 24, and 32. The $C_v$ peak for the KT transition is
characterized by a finite peak height in the limit of infinite size,
which is consistent with Fig.~\ref{fig:2dmcCv}.
We then compute the in-plane susceptibility $\chi$ defined as~\cite{cuccoli95}
(see Fig.~\ref{fig:2dmcChi})
\begin{equation} \label{eq:chixy}
\chi \equiv (\chi_x + \chi_y)/2 ,
\end{equation}
where the susceptibility in the $\alpha$ direction ($\alpha = x, y$)
is written as
\begin{equation} \label{eq:chi}
\chi_\alpha = \frac{1}{N}\left\langle \left( \sum_i s_\alpha^i \right)^2 \right\rangle.
\end{equation}
The in-plane susceptibility $\chi$ can be used to determine the
critical temperature $T_c$: We use 
the relation $\chi \sim L^{2-\eta}$ and the condition
that the exponent $\eta$ has value 1/4 at $T_c$, and obtain
$T_c = 0.621(3)$ (see the inset of Fig.~\ref{fig:2dmcChi}).
It is again to be noted that $T_c= 0.621(3)$ obtained here for 
the 2D CHS$XY$ model in the spherical coordinate representation
is different from $T_c = 0.699(3)$ obtained in Ref.~\onlinecite{cuccoli95}
for the conventional CHS$XY$ model.

\section{Relaxational Dynamics Simulation}
\label{sec:rd}

In the relaxational dynamics simulations in both 3D and 2D, we use the
second-order Runge-Kutta-Helfand-Greenside (RKHG) algorithm~\cite{batrouni85}
and the equations of motion in Eq.~(\ref{eq:rd}) are integrated numerically 
with the discrete time step $\Delta t = 0.05$.
The relaxational dynamics with the representation $\{\theta, \phi \}$ 
used here (see Sec.~\ref{sec:model}) is more convenient than the representation
$\{s^z, \phi \}$ since no constraint on $\theta_i$ and $\phi_i$
is required while the constraint $|s_i^z| \leq 1$ should be explicitly fulfilled
in the latter representation.
After neglecting initial transient behaviors, ensemble averages
of static physical quantities can be computed from the time averages
of those quantities due to the ergodicity of the system.
In contrast to the spin-dynamics 
method,~\cite{SD2D,skoh,krech99} where
initial configuration for dynamic calculation should be 
generated from the MC simulation, one can in relaxational
dynamics take any initial configuration to start with; as time proceeds, the dynamics
intrinsically generates equilibrium fluctuations.

\subsection{Three-dimensional lattice} \label{subsec:3drd}
We first present the static results in 3D. 
Figure~\ref{fig:3drdB} shows
the determination of $T_c$ from Binder's 
cumulant [see Eqs.~(\ref{eq:Binder}) and (\ref{eq:Uscale})]: the crossing point
gives $T_c=1.245(2)$ and $U^* = 0.585(3)$. In dynamic simulations
the inevitable finite time step $\Delta t$ causes an effective
shift of the temperature (see Ref.~\onlinecite{bjkim99a} for discussions).
Of course in the limit of $\Delta t \rightarrow 0$ this temperature shift 
vanishes~\cite{bjkim99a} and the critical temperatures determined
from dynamic simulation and MC simulation become identical.
The effective temperature shift in the RKHG algorithm used here
is much smaller than the simple Euler algorithm, and the deviation 
in $T_c$ in the current 3D case is less than 1\%.
Within numerical accuracy we find that $U^*$ computed from relaxational
dynamics (see Fig.~\ref{fig:3drdB}) agrees with that from MC in 
Sec.~\ref{subsec:3dmc}, and thus also with values in 
Refs.~\onlinecite{nho99} and \onlinecite{krech99}.
We display the determination of the critical exponent $\nu$
for the relaxational dynamics in the inset of Fig.~\ref{fig:3drdB},
where $\nu = 0.66(6)$ is obtained (compare with Fig.~\ref{fig:3dmcB}
for MC in 3D).
The values of $\nu$ and $U^*$ found from the relaxational dynamics
of the 3D CHS$XY$ model in the spherical coordinate representation
again confirm that the static universality class
of the model is identical to the usual 3D $XY$ universality class.

We next investigate the dynamic critical behaviors.
One convenient way of characterizing the dynamic universality class
is to compute the total spin time correlation function $G(t)$
defined as~\cite{skoh}
\begin{equation} \label{eq:G}
G(t) \equiv \frac{\langle S_x(t) S_x(0) + S_y(t) S_y(0) \rangle}
            {\langle S_x^2(0) + S_y^2(0)\rangle}, 
\end{equation}
where the total spin vector ${\bf S} = (S_x, S_y, S_z)$ 
is given in Eq.~(\ref{eq:totalS})
and the $\langle \cdots \rangle$ is substituted by the time
average in the relaxational dynamics study.
Since $G(t = 0) = 1$ at any $T$ and $L$, the finite-size scaling
of $G(t)$ is written in a very simple form,
\begin{equation} \label{eq:Gscale}
G(t,L,T) = g\bigl( tL^{-z}, [T-T_c]L^{1/\nu} \bigr), 
\end{equation}
where the first scaling variable is the ratio between
the time $t$ and the characteristic time scale $\tau \sim L^z$
with the dynamic critical exponent $z$,
and the second scaling variable comes from the ratio
between the system size $L$ and the coherence length 
$\xi \sim (T-T_c)^{-\nu}$ with the static critical exponent $\nu$.

At $T=T_c$, the above scaling form reduces to a simpler
form with a single scaling variable
\begin{equation}
G(t,L,T_c) = g(tL^{-z},0), 
\end{equation}
and all $G$'s at different system sizes should collapse
to a single curve once the correct value of $z$ is chosen.
In Fig.~\ref{fig:3drdGTc} (see also Figs.~\ref{fig:3drdGint} and \ref{fig:3drdGall}), 
$G$ at $T=1.25$ is shown 
(a) as a function of the time $t$ and (b) as a function
of the scaling variable $tL^{-z}$ with $z=2.0$.
All curves at different sizes $L=4$, 6, 8, 10, and 12 are shown
to collapse relatively well to a single curve with the dynamic
critical exponent $z=2.0$, although the quality of the
collapse is not perfect: This is because 
$T=1.25$ and $z=2.0$ chosen in Fig.~\ref{fig:3drdGTc} can be
slightly different from the true $T_c$ and $z$ (see below
for a more precise determination).

The other implication of the above finite-size scaling form
with two scaling variables in Eq.~(\ref{eq:Gscale}) is that 
if the first scaling variable is fixed to a certain constant
value $a \equiv tL^{-z} = {\rm const}$, it again reduces to a simple
form:
\begin{equation}
G(t,L,T) = g\bigl( a, [T-T_c]L^{1/\nu} \bigr).
\end{equation}
It then suggests that when $G$'s with fixed $a$ 
at different system sizes
are plotted as functions of $T$, all curves should cross
at a single point at $T=T_c$ if the correct value
of $z$ is chosen; this provides an independent method to 
determine $T_c$ (and $z$ at the same time).
Figure~\ref{fig:3drdGint} displays this intersection plot
with $z=2.05$ and $a = 0.53$ (this value of $a$, with which 
the intersection occurs at $G \approx 0.5$, is taken
only as an example; in a broad range of $a$ the similar
intersection plot is achieved).
We try different values of $z$ and $a$, and it is
concluded that $z=2.05(5)$ and $T_c=1.245(3)$,
the latter of which is in an agreement with 
the previously determined value from Binder's
cumulant in Fig.~\ref{fig:3drdB}. 

In summary of this section,
the relaxational dynamics study applied for the 3D CHS$XY$ model 
in the spherical coordinate representation has revealed that
this model belongs to the 3D $XY$ static universality class
characterized by $\nu \approx 0.67$ and $U^* \approx 0.586$,
while the dynamic critical exponent has the value $z \approx 2.0$.
We note that this value $z\approx 2$ is in accord with the
model A in the Hohenberg-Halperin classification~\cite{hh}
as well as with $z\approx 2.015$ found from the dynamic
renormalization-group calculation in Ref.~\onlinecite{wickham00}.
On the other hand, many studies on the 3D $XY$ model
with the resistively shunted junction dynamics,~\cite{larsB,lars3D,khlee92}
the relaxational dynamics~\cite{larsB,lars3D} under the fluctuating-twist 
boundary condition,~\cite{bjkim99a} and the MC dynamics 
for both phase~\cite{bjkim3dmc} and vortex~\cite{LCG3D} representations have 
observed $z\approx 1.5$. Also, the spin dynamics for the conventional
3D CHS$XY$ model also has yielded $z$'s that are significantly
different from the value 2: $z \approx 1.5$ in Ref.~\onlinecite{skoh},
and $z_x \approx 1.38(5)$ and $z_z = 1.62(5)$ in Ref.~\onlinecite{krech99}

\subsection{Two-dimensional lattice} \label{subsec:2drd}
\begin{figure}
\centering{\resizebox*{!}{6.0cm}{\includegraphics{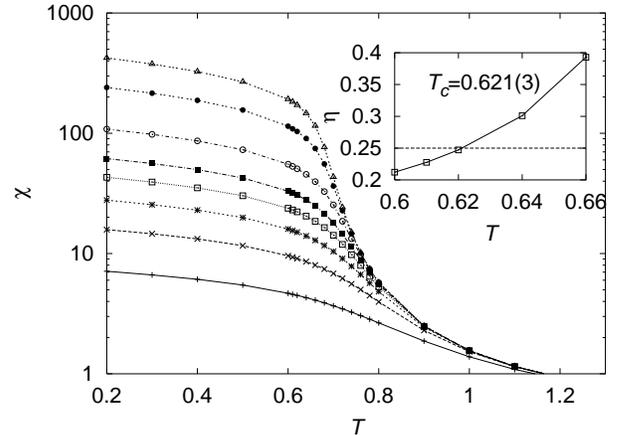}}}
\caption{
In-plane susceptibility $\chi$ from MC simulation in 2D
versus temperature $T$ at system sizes $L=4$, 6, 8, 10, 12, 16, 24, and 32 
(from bottom to top).
Inset:
The exponent $\eta$, obtained from $\chi(T,L) \sim L^{2-\eta(T)}$
is shown as a function of $T$.  From the condition $\eta(T_c) = 1/4$,
$T_c = 0.621(3)$ is found.
}
\label{fig:2dmcChi}
\end{figure}

\begin{figure}
\centering{\resizebox*{!}{6.0cm}{\includegraphics{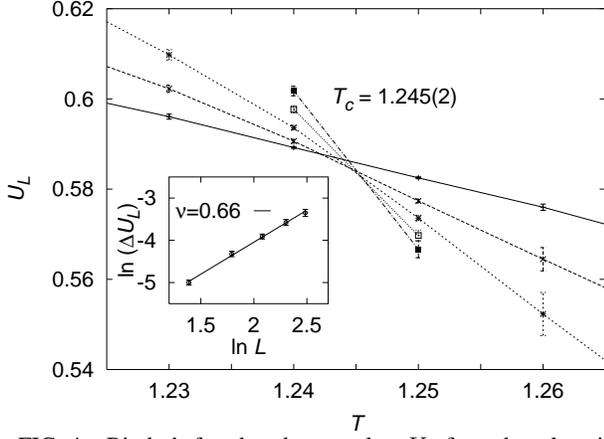}}}
\caption{
Binder's fourth order cumulant $U_L$ from the relaxational dynamics simulation in 3D as a function of $T$
for $L=4$, 6, 8, 10, and 12 (from top to bottom on the right-hand side
of the crossing point). 
The crossing point gives the estimation $T_c = 1.245(2)$.
Inset: Determination of $\nu$ 
from the least-square fit; $\nu = 0.66(6)$ is obtained
(to be compared with Fig.~\protect\ref{fig:3dmcB} for MC).
}
\label{fig:3drdB}
\end{figure}

\begin{figure}
\centering{\resizebox*{!}{6.0cm}{\includegraphics{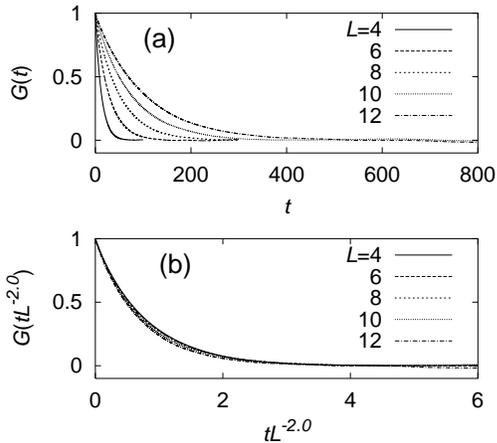}}}
\caption{
Total spin correlation function $G$ from relaxational dynamics in 3D
at $T=1.25$ is shown for $L=4$, 6, 8, 10, and 12, 
as a function of (a) the time $t$ and (b) the 
scaling variable $tL^{-z}$ with $z=2.0$.
The curve collapse in (b) implies that $T_c \approx 1.25$ and
$z \approx 2.0$. See Figs.~\protect\ref{fig:3drdGint} and 
\protect\ref{fig:3drdGall} for more precise determination
of $T_c$ and $z$.
}
\label{fig:3drdGTc}
\end{figure}

\begin{figure}
\centering{\resizebox*{!}{6.0cm}{\includegraphics{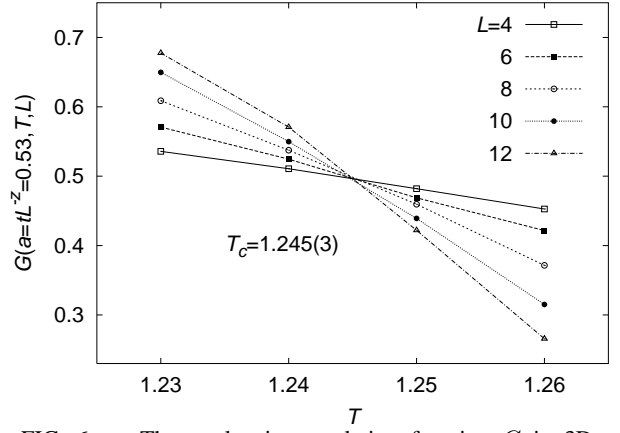}}}
\caption{
The total spin correlation function $G$ in 3D with $a \equiv tL^{-z} = 0.53$ 
($z=2.05$) is shown as a function of temperature $T$ for various system 
sizes $L=4$, 6, 8, 10, and 12. 
All curves cross at $T \approx 1.245$,
which is in a very good agreement with $T_c = 1.245(2)$ found in static
calculation with relaxational dynamics (see Fig.~\protect\ref{fig:3drdB}).
The value of $a = 0.53$ is chosen only for convenience ($a=0.53$
makes the intersection occur at $G \approx 0.5$); in a broad
range of $a$, the quality of this intersection plot is very good
if $z=2.05$ is chosen. The other values of $z$
and $a$ are tried and $z=2.05(5)$ and $T_c = 1.245(3)$ are concluded.
}
\label{fig:3drdGint}
\end{figure}

\begin{figure}
\centering{\resizebox*{!}{6.0cm}{\includegraphics{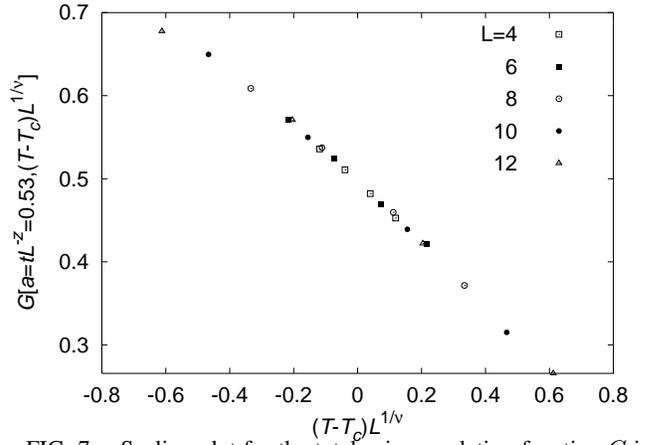}}}
\caption{
Scaling plot for the total spin correlation function $G$ in 3D with
$a \equiv tL^{-z} = 0.53$ and $z=2.05$.
All data points in Fig.~\protect\ref{fig:3drdGint} collapse to
a smooth curve with $T_c = 1.245$ and $\nu = 0.67$. The quality of
the curve collapse at different $a$ values are similar to this one
in a broad range of $a$.
}
\label{fig:3drdGall}
\end{figure}

\begin{figure}
\centering{\resizebox*{!}{6.0cm}{\includegraphics{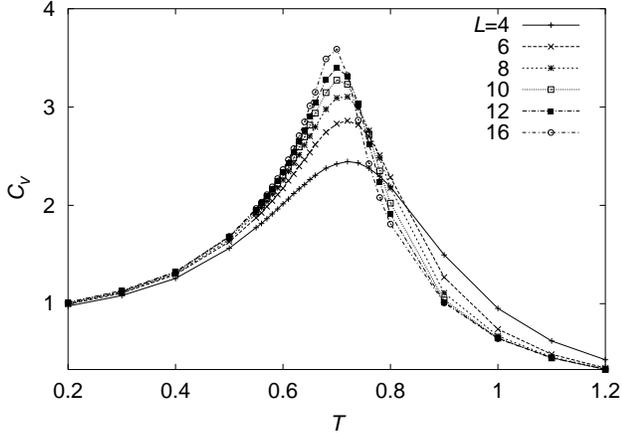}}}
\caption{
The specific heat $C_v$ versus $T$ from relaxational dynamics simulation
in 2D for various system sizes $L=4$, 6, 8, 10, 12, and 16 (from bottom
to top).
(Compare with Fig.~\protect\ref{fig:2dmcCv} that has been
obtained from independent MC simulations.)
}
\label{fig:2drdCv}
\end{figure}

\begin{figure}
\centering{\resizebox*{!}{6.0cm}{\includegraphics{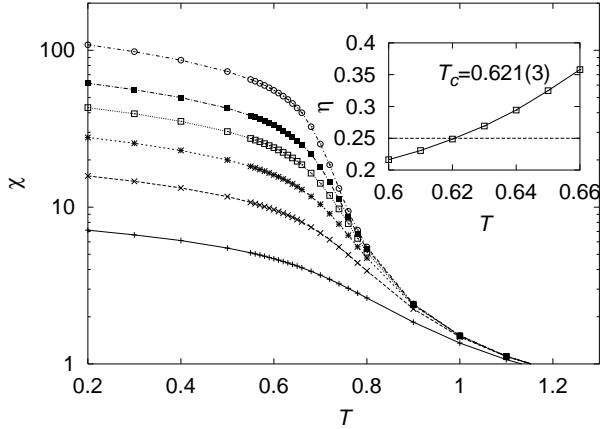}}}
\caption{
In-plane susceptibility $\chi$ from relaxational dynamic simulation in 2D
versus temperature $T$ for various system sizes $L=4$, 6, 8, 10, 12, and 16.
As expected, the relaxational dynamics simulations give quantitatively
similar curves (compare with Fig.~\protect\ref{fig:2dmcChi}).
Inset: The exponent $\eta$ as a function of $T$ is displayed
and $T_c = 0.621(3)$ is found.
(Compare with Fig.~\protect\ref{fig:2dmcChi} for MC.)
}
\label{fig:2drdChi}
\end{figure}

The static results obtained from the time averages during the numerical
integrations of the relaxational dynamic equations of 
motion~(\ref{eq:rd}) in 2D are first presented. 
The specific heat $C_v =  (\langle H^2 \rangle - \langle H \rangle^2)/T^2N$
with $N=L^2$, and the in-plane susceptibility $\chi$
in Eqs.~(\ref{eq:chixy}) and (\ref{eq:chi})
are exhibited as functions of the temperature $T$ in Figs.~\ref{fig:2drdCv}
and \ref{fig:2drdChi}, respectively. As expected, the static calculations
from the relaxational dynamics simulations result in basically the same
results as from the MC simulations in Figs.~\ref{fig:2dmcCv} and \ref{fig:2dmcChi}
in Sec.~\ref{subsec:2dmc}. From $\chi$ in Fig.~\ref{fig:2drdChi}, 
we locate $T_c$ in the inset of Fig.~\ref{fig:2drdChi} in the same way 
as in Fig.~\ref{fig:2dmcChi}
in Sec.~\ref{subsec:2dmc}, through the use of $\chi \sim L^{2-\eta}$ with
$\eta = 1/4$ at $T_c$. In 2D, we find $T_c = 0.621(3)$, which is identical
to $T_c$ found from MC simulation in Fig.~\ref{fig:2dmcChi} within numerical errors.

\begin{figure}
\centering{\resizebox*{!}{6.0cm}{\includegraphics{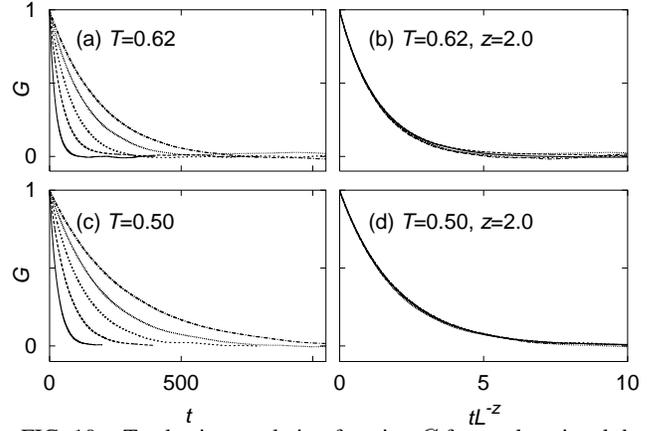}}}
\caption{
Total spin correlation function $G$ from relaxational dynamic
simulation in 2D at (a) $T=0.62\approx T_c$ and
(c) $T=0.50$ as a function of the time $t$ for $L=4$, 6, 8, 10, and 
12 (from left to right);
(b) and (d) display the corresponding scaling plots
with the scaling variable $tL^{-z}$ and
$z \approx 2.0$ is found at both $T=0.62$ and $0.50$, 
implying that $z(T) \approx 2.0$ in the whole low-temperature 
phase.
}
\label{fig:2drdG}
\end{figure}

We next turn to the investigation of the dynamic universality class 
of the 2D CHS$XY$ model in the spherical coordinate representation
under the relaxational dynamics.
In general, the 2D systems with the KT transition
are quasicritical in the whole low-temperature phase.
This means that when $T \leq T_c$ we cannot use the finite-size scaling form
in Eq.~(\ref{eq:Gscale}) since the coherence length $\xi$ is infinite.
In 2D, we then write the scaling form for $T\leq T_c$ as follows:
\begin{equation} \label{eq:2dGscale}
G(t,T,L) = g(tL^{-z(T)}),
\end{equation}
where the dynamic critical exponent $z(T)$ is allowed to vary
with temperature. (More precisely, the scaling function $g$ should also
depend on $T$.)
Figure~\ref{fig:2drdG}(a) displays the total spin time correlation function 
$G$ in Eq.~(\ref{eq:G}) at $T=0.62$  as a function of the time $t$ for various system
sizes $L=4$, 6, 8, 10, and 12. If we put $z = 2.0$ in Eq.~(\ref{eq:2dGscale})
all curves in Fig.~\ref{fig:2drdG}(a) collapse to a single curve
as shown in Fig.~\ref{fig:2drdG}(b). Consequently, we conclude that
the 2D CHS$XY$ model in the spherical-coordinate representation
under relaxational dynamics has the dynamic critical
exponent $z \approx 2.0$ at $T_c$.
We show in Fig.~\ref{fig:2drdG}(c) and (d) the similar scaling plot at $T=0.50$,
which is significantly lower than $T_c$. Interestingly, we again find
$z\approx 2.0$ at $T=0.50$, which suggests that this model
has $z \approx 2.0$ in the whole low-temperature phase.

In many existing analytical and simulational studies of the 2D $XY$ model 
in its original form, $z(T)$ has been found to have value 2 at $T_c$, 
and to increase as $T$ is 
decreased.~\cite{minnhagen:rmp,ahns,bjkim01,bjkim99a,larsB,holmlund96,weber2D}
Since $z(T)$ can also be related with the nonlinear $IV$ exponent $a$
by $a(T) = z(T) + 1$, which is usually measured in experiments,
there are also experimental papers with the same conclusion.~\cite{foot:2dexp}
In contrast, there exist studies with other conclusions: 
For example, in Ref.~\onlinecite{shortXY}
the decay from nonequilibrium to equilibrium (this technique
is often called ``short-time relaxation method'') in the MC dynamics
has been found to result in $z(T) \approx 2$ at any $T$ below $T_c$, 
and the same has been concluded in Ref.~\onlinecite{larsB} from
the similar short time relaxation method but applied for the relaxational
dynamics of the $XY$ model.
Also in Ref.~\onlinecite{skkim}, the scaling of the
total spin correlation function has been investigated
in the same way as in the present paper, and $z \approx 2$
in the whole low-temperature phase has been concluded
for the relaxational dynamics of the 2D $XY$ model.

The spin dynamics study of the conventional 2D CHS$XY$ model
in Ref.~\onlinecite{SD2D} has obtained
$z \approx 1.0$, which is close to $z = d/2$ ($d=2$ in 2D)
for the model E value in Hohenberg-Halperin classification.~\cite{hh}
While the models in Ref.~\onlinecite{SD2D} and in the present work belong
to the same static universality class, they do not need to belong to the same dynamic
universality class: In spin dynamics, the $z$ component $S_z$ of
the total spin  is a constant of motion since the 
Hamiltonian $H$ commutes with the spin operator in the 
$z$ direction. On the other hand,
the relaxational dynamics is not based on this commutation
relation and $S_z$ is not a conserved quantity.

Although 2D and 3D CHS$XY$ model have the same $z$, 
it should be kept in mind that their critical behavior
is completely different: In 2D, the whole low-temperature
phase is quasicritical and one can associate $z(T)$
at each temperature to make $G$ at different
sizes collapse to a single curve as displayed in Fig.~\ref{fig:2drdG}.
In 3D, on the other hand, the system is critical 
only at $T_c$ and the curve collapse with the single scaling
variable $tL^{-z}$ as shown in Fig.~\ref{fig:3drdGTc}
is not found at any other temperatures.

\section{Summary}
\label{sec:sum}
We have investigated the static and the dynamic
universality class of the two- and three-dimensional
CHS$XY$ model where three-dimensional
classical spins interact with each other through the Hamiltonian
with only in-plane components coupled.

The spherical polar $\theta$ and azimuthal $\phi$ angles of 
the spin direction, both with uniform measures in phase space,
are taken as dynamic variables, which
leads to the simple relaxational dynamic equations of motion
since the constraint $|{\bf s}_i| = 1$ is fulfilled 
automatically. 
It is to be noted that the relaxational dynamics method
makes it possible to study both the dynamic and the static
properties on the same footing, in contrast to the spin dynamics method.
In other words, in the spin dynamics method
initial configurations for dynamic calculation should be
generated from the MC simulation, while one can in relaxational
dynamics take any initial configuration to start with; 
the relaxational dynamics intrinsically generates 
equilibrium fluctuations as the system evolves in time.

From the static calculations based on the relaxational dynamics method,
it was explicitly verified 
that both the 2D and 3D CHS$XY$ models in the spherical-coordinate representations
belong to the expected 2D and 3D $XY$ static universality classes, respectively.
The dynamic critical exponent $z$ has been found to
be different from values obtained from various other
existing studies of the $XY$ model including relaxational dynamics.
The value $z \approx 2.0$ found here for both 2D and 3D
implies that the relaxational dynamics of the CHS$XY$ model
is governed by the model A description in the Hohenberg-Halperin
classification.~\cite{hh} 

\acknowledgments
B.J.K. is grateful for the hospitality during his visit at 
Chungbuk National University, where this work was begun.
This work was supported by the Swedish Natural Research Council
through Contract No. F 5102-659/2001 (B.J.K. and P.M.), and
S.K.O. and J.S.C. were supported by the Korea Research Foundation
through Grant KRF-99-005-D00034.
 

\begin{thebibliography}{10}
\bibitem{foot:def}
Throughout this paper, the CHS$XY$ model represented by 
the Hamiltonian~(\ref{eq:Hzphi}) with $s^z$ and $\phi$ subject
to uniform measure in phase space is referred as ``the conventional
CHS$XY$ model'', while we call the model~(\ref{eq:Htp}) with the
uniform measures for $\theta$ and $\phi$ ``the CHS$XY$ model in
the spherical coordinate representation'', for convenience.
\end{thebibliography}

\narrowtext

\vskip 6cm

\newpage

\vskip 10cm

\end{multicols}
\end{document}